# Enhancing Elasticity of SaaS Applications using Queuing Theory

Ashraf A. Shahin[1,2]

[1]College of Computer and Information Sciences,
Al Imam Mohammad Ibn Saud Islamic University (IMSIU)
Riyadh, Kingdom of Saudi Arabia

[2]Department of Computer and Information Sciences, Institute of Statistical Studies & Research,
Cairo University,
Cairo, Egypt

*Abstract*—Elasticity is one of key features of cloud computing. Elasticity allows Software as a Service (SaaS) applications' provider to reduce cost of running applications. In large SaaS applications that are developed using service-oriented architecture model, each service is deployed in a separated virtual machine and may use one or more services to complete its task. Although, scaling service independently from its required services propagates scaling problem to other services, most of current elasticity approaches do not consider functional dependencies between services, which increases the probability of violating service level agreement. In this paper, architecture of SaaS application is modeled as multi-class *M/M/m* processor sharing queuing model with deadline to take into account functional dependencies between services during estimating required scaling resources. Experimental results show effectiveness of the proposed model in estimating required resources during scaling virtual resources.

## I. Introduction

In the last few years, Software as a Service (SaaS) has rapidly spread in many areas. SaaS is a software delivery model in which software is delivered to customers as a service [1]. Instead of delivering individual application instance for each tenant, one application instance serves thousands of tenants [2]. Nowadays, several SaaS companies, such as Salesfore.com, NetSuite, and Success Factors, utilize elasticity feature of cloud computing to ensure lowest cost of service delivery. However, developing multi-tenant SaaS application to serve thousands of tenants with thousands of users for each tenant is a very hard and expensive task due to large number of factors that have to be considered during development phases, such as customizability, security, scalability, and pricing.

Most of current SaaS applications have been developed using service-oriented architecture (SOA) model [1]. In SOA model, each application is a collection of services that are organized in several layers. Each service uses services in the lower layer to complete its tasks. In large SaaS applications, each service is deployed in a separated virtual machine. Although, one of primitive assumptions is that scaling any service has to be reflected in all required services, most of current researches do not consider functional dependencies between services and scale them separately. As consequence, scaling problems are shifted from layer to next layer. Unfortunately, the problem is not only specifying functional dependencies between services but also specifying number of virtual machines that have to be added or removed.

For example, suppose we have three services *X*, *Y*, and *Z*. Service *X* uses services *Y* and *Z* to complete its tasks. Service *X* receives three types of requests *A*, *B*, and *C*. Service *X* uses service *Y* to complete requests of type *A*, uses service *Z* to complete requests of type *B*, and uses service *Y* and service *Z* to complete request of type *C*. If service *X* is detected as overloaded, scaling service *X* independently from *Y* and *Z* moves overloading problem to *Y*, *Z*, or both of them. However, which service has to be scaled and what is the optimal number of VMs instances that have to be added to or removed from each service? This depends on types of arriving requests. If overloading is occurred due to high number of requests of type *A*, then adding more VMs to service *Z* will waste resources and reduce revenue. Collecting such information without modeling functional dependencies is a very hard task.

Thus, this paper models SaaS applications as multi-class *M/M/m* processor sharing queuing model with deadline to consider functional dependencies and requests' types during estimating required scaling resources. The proposed model reflects scaling actions on many metrics such as CPU utilization, response time, and throughput, which are commonly used by most of current auto-scaling techniques to trigger auto-scaling actions. Therefore, SaaS application providers can apply the proposed model with any auto-scaling technique to put into account functional dependencies between services.

Queuing network models have been extensively applied in many areas and have proven their efficiency in representing and analyzing resource-sharing systems such as computer systems [3]. According to Kendall's Notation, the first *M* in *M/M/m* queuing model represents arrival process, which is Markov arrival process. It has been theoretically proved that if large number of customers make independent decisions of when to request service, the resulting arrival process will be Markov arrival process [4]. The second *M* in *M/M/m* queuing model represents service process, which is Markov service process. Third *m* represents number of parallel servers that provide one service. Servers receive requests from different classes and serve them according to processor sharing discipline.

Effectiveness of the proposed model has been evaluated by comparing performance of auto-scaling algorithms with and without the proposed model. Simulation results show that the proposed model reduces violation of Service Level Agreement and increases revenue.

The rest of this paper is organized as follows. Section 2 describes the related work. Section 3 briefly describes the proposed model. Section 4 experimentally demonstrates the effectiveness of the proposed model. Finally, Section 6 concludes.

## II. RELATED WORK

Although, several auto-scaling approaches have been proposed in the last few years [6, 7, 8, 9, 10], most of current auto-scaling approaches do not consider functional dependencies between application's services. Current auto-scaling approaches can be categorized into two main categories: reactive and proactive approaches. Reactive auto-scaling approaches scale computational resources based on some rules and according to some metrics such as memory utilization, CPU utilization, throughput, and response time [15, 16, 17, 18]. However, relations between metrics of related services are not modeled. Therefore, impact of scaling service is unknown until its occurrence.

In another hand, proactive auto-scaling approaches trigger auto-scaling operations based on predicted workload. Different time series techniques such as Support Vector Machine, Exponential Smoothing, and Neural Networks have been used in predicting future workload [13, 14, 17, 19, 20]. Although, functional dependencies between application's services are very effective factors in predicting future workload, most of current proactive techniques do not consider it. This section overviews some of current approaches.

Biswas, et al. [5, 21] have proposed framework to provide virtual private cloud for a single client enterprise. Proactive auto-scaling technique has been proposed to provision and release resources from public cloud according to predicted system load. Support vector machine and linear regression have been employed to predict future load. In [6] Biswas, et al. have proposed a reactive auto-scaling algorithm to serve incoming requests with considering their service level agreements. The proposed algorithm scales resources based on profit that is gained from serving incoming requests and based on cost benefit to the user.

Sellami et al. [7, 8] have proposed threshold based auto-scaling approach to offer dynamic service instances for multi-tenant business processes. The proposed approach considers functional dependencies between each multi-tenant process and its services during deciding scaling action. The proposed approach has been encapsulated into middleware layer between software and platform layers.

Xiao et at. [9] have modeled automatic scaling problem as Class Constrained Bin Packing problem where each server is a bin and each class represents an application. To scale provisioned resources, semi-online color set algorithm has been proposed. However, they have encapsulated each application instance inside a virtual machine (VM), which is not applicable in large applications.

Ahn et al. [10] have proposed auto-scaling method to support execution deadline. The proposed method can handle Bag-of-Tasks jobs and workflow jobs. Jobs in Bag-of-Tasks can be scheduled separately from each other while jobs in workflow have to be scheduled in order of its dependency. The proposed method has been evaluated by using Cloudsim, which shows that the proposed auto-scaling method increases resources utilization.

Chaloemwat et al. [11] have tried to enhance performance of threshold-based auto-scaling techniques by using Skewness algorithm and VMs migration. The effectiveness of the proposed enhancement has been proven by comparing performance of threshold-based auto-scaling techniques with and without the proposed enhancement.

Srirama et al. [12] has proposed resource provisioning policy that takes into account lifetime, periodic cost and configuration cost of each instance type to find most optimal combination of possible instance types. The auto-scaling problem is represented as a linear programming model. Solution of this linear programming model will provides optimal number of VMs instances from each instances type that must be added or removed to achieve workload with minimum cost. Unfortunately, linear programming model can provide solutions for small number of VMs and cannot deal with large systems.

Hirashima et al. [13] have proposed threshold based auto-scaling mechanism that proactively adjusts resource to fulfill incoming workload based on predicted workload. Autoregressive Integrated Moving Average model has been exploited to forecast future workload. Moreover, the proposed mechanism reactively adapts virtual resources if unpredictable workload arrives. However, performance of the proposed mechanism has not been evaluated with unpredictable workload.

Khatua et al. [14] have proposed threshold based auto-scaling algorithm that adopts virtual resources proactively according to predicted workload. The proposed algorithm predicts workload by using Auto-regressive Integrated Moving Average (ARIMA) model.

Nikravesh et al. [22] have proposed auto-scaling system, which predict workload using two time-series prediction algorithms: Support Vector Machine (SVM) and Neural Networks (NN). The proposed system automatically switches between SVM and NN based in patterns of workload. SVM is used with periodic workload patterns while NN is used with unpredicted workload pattern. Although, functional dependency is an important factor in predicting workload, functional dependency has not been considered during predicting future workload.

Liao et al. [23] have proposed dynamic threshold based auto-scaling strategy for Amazon web services. The proposed strategy adapts thresholds according to demand for resources. Upper threshold is set in the range 50%–75% and lower threshold is set to the range 5%–30%. Upper and lower thresholds are adapted proportionally with expansion process of VMs.

Tang et al. [24] have proposed reinforcement learning based auto-scaling algorithm. Workload is categorized into normal workload (daily busy-and-idle workload) and burst workload. Auto-scaling problem is model as Markov Decision Process (MDP) model and Reinforcement Learning is applied to decide time to scale up or down and to decide number of VM instances to be added or removed.

Chen et al. [25] have proposed hybrid auto-scaling mechanism. The proposed mechanism predicts next CPU usage rate based on historical data by applying several time series techniques such as Autoregressive–Moving-Average model, Autoregressive model, Exponential Smoothing model, Moving Average model, and Naïve model. The proposed mechanism reactively scales resources to minimize affects of wrong workload prediction.

### III. SaaS Application Model

This paper deals with SaaS applications that cannot be encapsulated in one VM and are developed using Service-Oriented Architecture model. Each service is deployed in a separated VM instance and can be scaled up or down by adding or removing VM instances. Each VM has a fixed processing capacity, which is divided into equal parts among all tasks (Processor Sharing (PS)). Thus, each task's service time depends on the total number of tasks that exist at the same time. No task can run simultaneously on more than one VM. Therefore, if number of tasks is less than number of VMs for a specific service, each task is processed by a single VM and the remaining VMs are idle. If number of tasks is greater than number of VMs, tasks are processed according to processor sharing discipline. In this paper, the term "web service" will be used to refer to service component in SaaS application.

Each web service receives requests from one or more upper web services and it can complete tasks by itself or by sending requests to lower web services. After receiving responses from lower web services, request will be completed and sent to upper web services as a response to its request. Web services receive requests from different types. Each type has its arrival rate, process rate, routing, and deadline. Requests from the same type are collected in a chain. A chain contains a set of classes to represent different processing phases for a specific type. Classes are distributed among different web services, and each request moves between these classes during it life.

For example, suppose we have a web service (node $M+1$) with $M$ upper web services (nodes $1, 2, .., M$) and $K$ lower web services (nodes $M+2, M+3, ..., M+K+1$) (see Fig. 1). According to processor sharing, if there are $N$ requests in node $M+1$ at time t, service time for these requests will be decreased by $1/N$ per unit of time. Total number of requests that are served in node $M+1$ at time $t$ is calculated as:

$$N = \sum_{r=1}^{R} n_r, \qquad (1)$$

where $n_r$ is the number of requests of class $r$ that are served in node $M+1$, $r = 1, 2, .., R$.

Node $M+1$ receives R classes of requests from upper web services and sends requests to lower web services synchronously or asynchronously. In Fig. 2, node $M+1$ sends asynchronous requests to nodes $M+2$ and $M+3$. Chain *1* describes routing behavior of type *1* requests. Request visits node *M+1* in class *a*, node *M+2* in class *b*, node *M+1* in class *c*, node *M+3* in class *d*, and node *M+1* in class *e*.

In some cases, node $M+1$ needs to use two or more nodes synchronously to complete specific request. In this case, several sub-requests are generated, processed in parallel, combined to one request, and sent back to node $M+1$. In Fig. 3, node $M+1$ sends synchronous requests to nodes $M+2$ and $M+3$. Fork node represents decomposition of request to two or more sub-requests, which will be processed in parallel by $M+2$ and $M+3$ nodes. Synchronizing node represents buffer that holds completed sub-requests until it can be recomposed with sub-requests from other sibling nodes. Join node represents recombination of completed sub-requests to one request again.

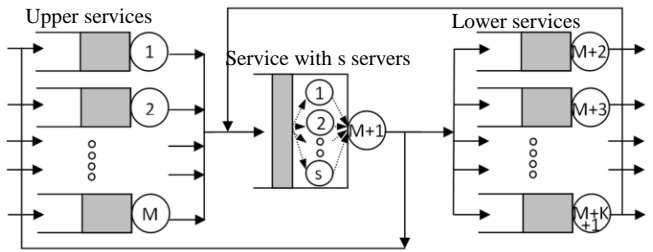

Fig. 1. Web service M+1 with its upper and lower web services

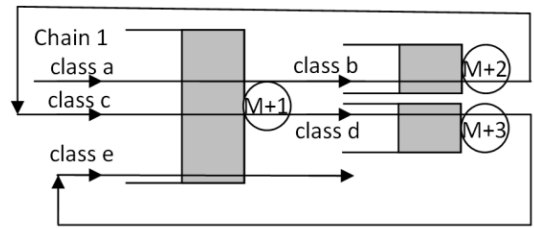

Fig. 2. Example of asynchronous requests from web service *M+1* to *M+2* and *M+3*

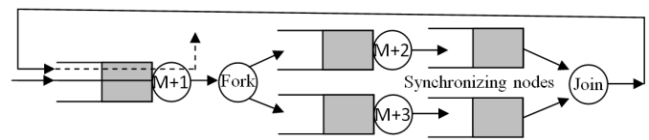

Fig. 3. Example of synchronous requests from web service *M+1* to *M+2* and *M+3*

In multiclass M/M/m processor sharing queuing systems, requests of class $r$ arrive to node $i$ according to Poisson process with rate $\lambda_{ir}$ and require service time $\mu_{ir}$ with exponential service process. Each class $r$ of requests has a deadline $d_r$. Arrival rates and service times are all assumed to be mutually independent. Deadline of each request class is specified according to required Service Level Agreement.

Request that is completed at node i will be sent to node from upper nodes (nodes *1, 2, .., M*), if it is completely finished. All nodes will receive responses from other services for their requests. Request will be sent to node from lower nodes (nodes $M + 2, M + 3, ..., M + K + 1$), if it still requires more processing. Request will be sent from node $i$ to node $i$ itself, if there is new program path. If deadline of any request expires, this request will exit the system, so that

$$\lambda_{ir} = \sum_{j,s} \lambda_{js}\, p_{j,s;i,r} \quad for\ j = 1,\dots,M+K+1 \quad (2)$$

where $p_{j,s;i,r}$ is the probability of sending requests from node $j$ of class $s$ to node $i$ of class $r$.

In root service, arrival rate of each request class is observable and can be measured easily. Probability $p_{j,s;i,r}$ can be specified by SaaS application providers based on business process workflow of their applications.

According to Burke's Theorem [26], the departure process from a M/M/m/∞ queue is Poisson, splitting a Poisson process randomly gives Poisson processes, and sum of Poisson processes is a Poisson process. Therefore, $\lambda_{ir}$ is Poisson.

In steady-state, total required service time from node $i$ at time $t$ is calculated as

$$\sum_{r=1}^{R} \lambda_{ir} \cdot \mu_{ir} \quad (3)$$

***Service time***: while arrival time and departure time of each request class are observable and can be measured easily, service time of each request class is not observable and cannot be measured easily (due to processor sharing). Therefore, service time $\mu_{ir}$ of requests of class $r$ that arrive to node $i$ can be calculated as following (with assuming homogeneity of servers)

$$\mu_{ir} = \sum_{t=t_0}^{t_d} \frac{S_i(t)}{N_i(t)} \quad (4)$$

where $t_0$ is observed arrival time of request of class $r$ to node $i$, $t_d$ is observed departure time of request of class $r$ from node $i$, $S_i(t)$ is number of running servers in node $i$ at time $t$, and $N_i(t)$ is total number of requests that are served in node $i$ at time $t$.

***Number of required servers***: processing sharing does not consider deadlines of request classes and gives the same amount of processing to all requests. Therefore, number of required servers at node $i$ to achieve incoming requests without violating Service Level Agreement is calculated as

$$s_i > \frac{\sum_{r=1}^{R} \lambda_{ir} \cdot \mu_{ir}}{MIN_{r=1}^{R} d_r} \quad (5)$$

where $s_i$ is the number of servers in node $i$, $MIN_{r=1}^{R} d_r$ is the minimum deadline of all request classes.

***Service rate***: with $s_i$ servers, node $i$ delivers service to requests of class $r$ at a rate of

$$\eta_i^r = \frac{s_i\, n_r\, \mu_{ir}}{N} \quad (6)$$

where $N$ is total number of requests that are served in node $i$ at time $t$. $n_r$ is number of requests of class $r$ in node $i$.

***Utilization***: utilization of node $i$ at time $t$, which is the fraction of time the servers in the node $i$ are busy, can be approximated to

$$U_i(t) = \frac{\sum_{r=1}^{R} \lambda_{ir} \cdot \mu_{ir}}{s_i} \quad (7)$$

***Throughput:***

Throughput $T_i^r$ of node $i$ from class $r$ at time $t$ is calculated as in [27]

$$T_i^r(t) = \eta_i^r\, U_i(t)$$

$$T_i^r(t) = \frac{s_i\, n_r\, \mu_{ir}}{N} \cdot \frac{\sum_{r=1}^{R} \lambda_{ir} \cdot \mu_{ir}}{s_i}$$

$$T_i^r(t) = \frac{n_r\, \mu_{ir}\, \sum_{r=1}^{R} \lambda_{ir} \cdot \mu_{ir}}{N} \quad (8)$$

Total throughput $T_i$ of node i is calculated as

$$T_i(t) = \sum_{r=1}^{R} T_i^r(t) \quad (9)$$

***Service size***: if the system is in steady-state ($\sum_{r=1}^{R}(\lambda_{ir}/s_i\mu_{ir}) < 1$), the probability of existing $(n_1, \dots, n_R)$ requests of classes $(1, \dots, R)$ can be calculated as in [27]

$$P(n_1, \dots, n_R) = \lim_{t \to \infty} P(X_1(t) = n_1, \dots, X_R(t) = n_R) \quad (10)$$

$$= G \cdot \left( N! / \left( \sum_{(n_1,\dots,n_R) \in \mathbb{N}^R} N!\, G \right) \right), where$$

$$G = \prod_{r=1}^{R} \frac{\rho_r^{n_r}}{n_r!},\ \rho_r = \lambda_{ir}/s_i\mu_{ir}$$

and $X_r(t)$ is the number of requests of class $r$ that are exist in the system at time $t$.

***Service capacity***: service capacity $\sigma_{i,(n_1,\dots,n_R)}$ is number of requests that can be accepted by node $i$, which already contains $(n_1, \dots, n_R)$ requests of classes $(1, \dots, R)$. $\sigma_{i,(n_1,\dots,n_R)} = (n'_1, \dots, n'_R)$ requests of classes $(1, \dots, R)$ if

$$\sum_{r=1}^{R} n'_r \cdot \mu_{ir} = s_i\, MIN_{r=1}^{R} d_r - \sum_{r=1}^{R} n'_r \cdot \mu_{ir} \quad (11)$$

***Response time***: for request with remaining service time $k$, the probability of departure after exactly $k+m$ time is represented as $P_{k,m}(n_1,..,n_R)$, which depends on number of requests $(n_1,..,n_R)$ of classes $(1,..,R)$ that exist in the system and depends on the remaining service time $k$.

$$P_{k,m}(n_1,..,n_R) = P(n_1,..,n_R)\ P_{k,m}(N) \quad (12)$$

where $P_{k,m}(N)$ is the probability of responding after exactly $k+m$ for request with remaining service time $k$ from node $i$, which contains $N$ requests. The probability $P_{k,m}(N)$ can be calculated by applying Random Quantum Allocation approximation model proposed by Braband in [28]. Request will leave the system immediately if its service time is finished. Therefore,

$$P_{0,m}(N) = \begin{cases} 1 & if\ m = 0 \\ 0 & if\ m \neq 0 \end{cases} \quad (13)$$

If remaining service time $k$ is greater than zero, the probability of responding after $k+m$ is calculated as following:

$$P_{k,m}(N) = \lambda q\ P_{k,m-1}(N+1) +$$

$$(1-\lambda q)\begin{pmatrix} \dfrac{\sigma_{N+1}}{N+1} \sum_{j=0}^{\sigma_{N+1}-1} \mu_j^{\sigma_{N+1}-1}\ P_{k-1,m}(N-j) \\ +\left(1-\dfrac{\sigma_{N+1}}{N+1}\right)\sum_{j=0}^{\sigma_{N+1}} \mu_j^{\sigma_{N+1}}\ P_{k,m-1}(N-j) \end{pmatrix},$$

$$N+1 > \sigma_{N+1},$$

$$\mu_j^N = \binom{N}{j}(\mu\ q)^j(1-\mu\ q)^{N-j} \quad (14)$$

$$P_{k,m}(-1) = P_{k,-1}(N) = 0$$

where $\lambda$ is average arrival rate. $q$ is time slice length, which is equal to time unit in this model. $\sigma_N$ is number of requests that can be accepted by node, which already contains $N$ requests. $\mu_j^N$ is probability $j$ requests leave node that contains $N$ requests. $\mu$ is average service time.

## IV. EVALUATION

To evaluate performance of the proposed model, threshold based auto-scaling algorithm (without workload prediction) proposed By Shahin in [29] has been implemented with and without the proposed model. Several web applications have been modeled using *Cloudsim* simulator with *NetworkCloudSim*. *NetworkCloudSim* is an extension of *CloudSim* to support modeling of generalized applications such as High Performance Computing (HPC), e-commerce, social network and web applications. For each application model, different chains have been defined and requests to each application are generated according to *ClarkNet* trace [30].

Fig. 4 shows model of sample application with 6 services. Each service has been deployed to a separated VM. During run time, number of running VMs in each service is ranged between 1 and 83 VMs. As shown in Table 1, 6 chains have been defined with 20 classes. Table 2 shows classes of each service. According to Table 1 and Table 2, the following probabilities are set to ones:

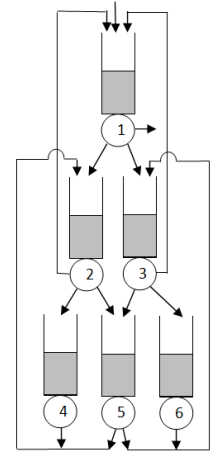

Fig. 4. Application model with three layers

TABLE I. CHAINS WITH REQUEST CLASSES

| Chain | Request Classes | Chain | Request Classes |
|---|---|---|---|
| $c_1$ | $r_1, r_2, r_3, r_4, r_5$ | $c_4$ | $r_{18}, r_{19}, r_{20}, r_{21}, r_{22}$ |
| $c_2$ | $r_6, r_7, r_8, r_9, r_{10}$ | $c_5$ | $r_{23}, r_{24}, r_{25}, r_{26}, r_{27}$ |
| $c_3$ | $r_{11}, r_{12}, r_{13}, r_{14}, r_{15}\ r_{16}, r_{17}$ | $c_6$ | $r_{28}, r_{29}, r_{30}, r_{31}, r_{32}\ r_{33}, r_{34}$ |

TABLE II. APPLICATION SERVICES WITH REQUEST CLASSES

| Service | Request Classes | Service | Request Classes |
|---|---|---|---|
| $s_1$ | $r_1, r_5, r_6, r_{10}, r_{11}, r_{17}$, $r_{18}, r_{22}, r_{23}, r_{27}, r_{28}, r_{34}$ | $s_4$ | $r_3, r_{13}$ |
| $s_2$ | $r_2, r_4, r_7, r_9, r_{12}, r_{14}, r_{16}$ | $s_5$ | $r_8, r_{15}, r_{20}, r_{30}$ |
| $s_3$ | $r_{19}, r_{21}, r_{24}, r_{26}, r_{29}, r_{31}, r_{33}$ | $s_6$ | $r_{25}, r_{32}$ |

$p_{s_1,r_1;s_2,r_2}, p_{s_1,r_6;s_2,r_7}, p_{s_1,r_{11};s_2,r_{12}}, p_{s_1,r_{18};s_3,r_{19}}, p_{s_1,r_{23};s_3,r_{24}},$

$p_{s_1,r_{28};s_3,r_{29}}, p_{s_2,r_2;s_4,r_3}, p_{s_2,r_4;s_1,r_5}, p_{s_2,r_7;s_5,r_8}, p_{s_2,r_9;s_1,r_{10}},$

$p_{s_2,r_{12};s_4,r_{13}}, p_{s_2,r_{14};s_5,r_{15}}, p_{s_2,r_{16};s_1,r_{17}}, p_{s_3,r_{19};s_5,r_{20}}, p_{s_3,r_{21};s_1,r_{22}},$

$p_{s_3,r_{24};s_6,r_{25}}, p_{s_3,r_{26};s_1,r_{27}}, p_{s_3,r_{29};s_5,r_{30}}, p_{s_3,r_{31};s_6,r_{32}}, p_{s_3,r_{33};s_1,r_{34}},$

$p_{s_4,r_3;s_2,r_4}, p_{s_4,r_{13};s_2,r_{14}}, p_{s_5,r_8;s_2,r_9}, p_{s_5,r_{15};s_2,r_{16}}, p_{s_5,r_{20};s_3,r_{21}},$

$p_{s_5,r_{30};s_3,r_{31}}, p_{s_6,r_{25};s_3,r_{26}}, p_{s_6,r_{32};s_3,r_{33}}$

Remaining probabilities are set to zeros.

As shown in Fig. 5 and Table 3, the proposed model improves number of completed requests, which reduces violation of Service Level Agreement and increases revenue. During run time, total number of running VMs is ranged between 6 and 415 VMs. By considering functional dependencies, VMs are added in advance to achieve incoming requests.

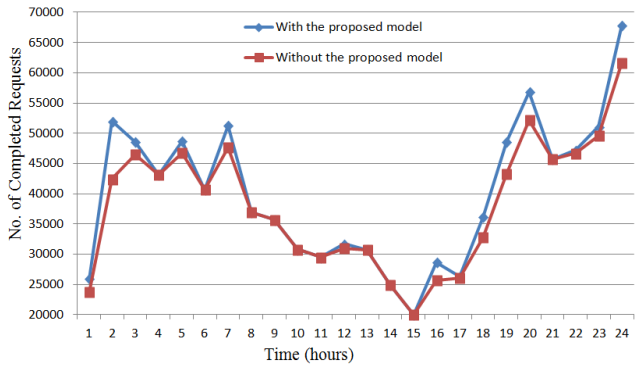

Fig. 5. Number of completed requests with and without the proposed model

TABLE III. NUMBER OF COMPLETED REQUESTS WITH AND WITHOUT THE PROPOSED MODEL

| Time (Hour) | With the proposed model | Without the proposed model | Time (Hour) | With the proposed model | Without the proposed model |
|---|---|---|---|---|---|
| 1 | 26000 | 23790 | 13 | 30700 | 30700 |
| 2 | 52000 | 42409 | 14 | 24900 | 24900 |
| 3 | 48600 | 46495 | 15 | 20000 | 20000 |
| 4 | 43100 | 43100 | 16 | 28700 | 25742 |
| 5 | 48700 | 46796 | 17 | 26300 | 26110 |
| 6 | 40700 | 40700 | 18 | 36200 | 32769 |
| 7 | 51300 | 47696 | 19 | 48600 | 43218 |
| 8 | 36900 | 36900 | 20 | 56800 | 52182 |
| 9 | 35700 | 35700 | 21 | 45700 | 45700 |
| 10 | 30800 | 30800 | 22 | 47200 | 46690 |
| 11 | 29500 | 29500 | 23 | 51100 | 49601 |
| 12 | 31700 | 30952 | 24 | 67900 | 61678 |

Implemented algorithm is a reactive algorithm. Consequently, it requires around 10 minutes to add new VM instances [30]. For example, if the first node is detected as over utilized due to large number of requests from *chain1*, without using the proposed model it will take around 30 minutes to be ready to response. This is due to adding VMs sequentially to nodes 1, 2, and 4. While, it will take around 10 minutes only if the proposed model is applied because VMs will be added to nodes 1, 2, and 4 concurrently. Therefore, the proposed model does not effect by number layers in applications. On the other hand, scaling without considering functional dependencies increases Service Level Agreement violation due to long sequence of scaling actions.

Fig. 6, Fig. 7, and Fig. 8 show number of completed requests by applications contain different numbers of layers. As shown in these figures, delays of scaling up applications that do not apply the proposed model are proportional to number of application layers.

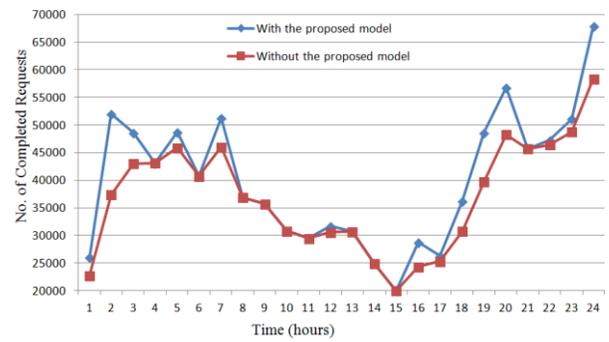

Fig. 6. Number of completed requests by application contains four layers

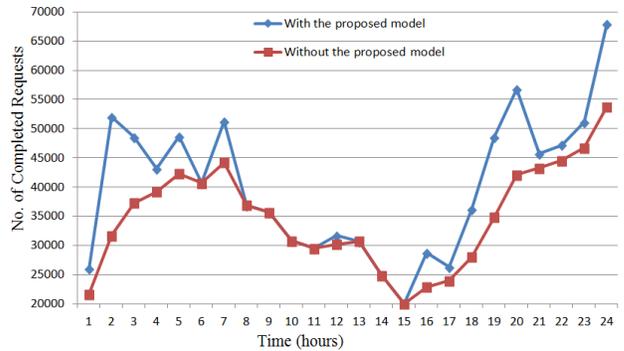

Fig. 7. Number of completed requests by application contains five layers

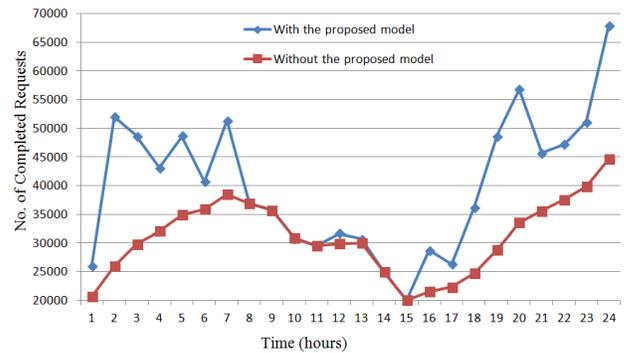

Fig. 8. Number of completed requests by application contains six layers

## V. CONCLUSION

Nowadays, several applications have been moved to cloud computing to benefit from its features. Cloud computing provides a large pool of resources that can be provisioned and release on demand. Some applications are small and can be encapsulated to a single VM. While large applications (such as social network) are distributed into several VMs. Although, functional dependency between services that are deployed to separated VMs has to be considered during application scaling, most of current scaling techniques do not consider functional dependency and scale services individually. This paper has modelled SaaS applications as multiclass *M/M/m* processor sharing queuing model with deadline to consider functional dependencies and requests' types during estimating required scaling resources. Based on experimental results, this paper concludes that modeling functional dependencies as multiclass *M/M/m* processor sharing queuing model improves performance of scaling algorithms.

In the future, the proposed model will be extended to include multiclass with different weights to represent different priorities that can be provided to customers.